\documentclass[twocolumn,nameyear]{autart}

\usepackage{graphicx}
\usepackage{amsmath}
\usepackage{amssymb}
\usepackage{xcolor}
\usepackage[authoryear,round]{natbib}

\usepackage{tikz}
\usetikzlibrary{shapes,arrows}

\newcommand{\RR}{{\mathbb R}}
\newcommand{\ZZ}{{\mathbb Z}}
\newcommand{\Ical}{{\mathcal I}}
\newcommand{\Pcal}{{\mathcal P}}
\newcommand{\Rcal}{{\mathcal R}}
\newcommand{\nom}{\mathrm{nom}}
\newcommand{\Res}{\operatorname{Res}}

\newtheorem{theorem}{Theorem}
\newtheorem{lemma}{Lemma}
\newtheorem{proposition}{Proposition}
\newtheorem{problem}{Problem}
\newtheorem{remark}{Remark}
\newenvironment{proof}[1][Proof]{%
    \noindent\textit{#1.}
}{    \hfill$\square$
}
\definecolor{myblue}{rgb}{0.540, 0.9, 1}

\begin{document}

\begin{frontmatter}

\title{Exact Characterization of Admissible Bandwidths for \\ Linear Disturbance Observers}

\thanks[cor]{Corresponding author}

\author[Purdue]{Hamin Chang}\ead{chan1232@purdue.edu}
\and
\author[Oxford]{Hyuntae Kim\thanksref{cor}}\ead{hyuntae.kim@eng.ox.ac.uk}            

\address[Purdue]{Elmore Family School of Electrical and Computer Engineering, Purdue University, West Lafayette, IN, USA}                     
\address[Oxford]{Department of Engineering Science, University of Oxford, Oxford, UK}

\begin{keyword}
Disturbance observer, Robust stability, Semialgebraic methods
\end{keyword}

\begin{abstract}        

This paper characterizes the entire set of admissible Q-filter bandwidths, equivalently parameterized by the time constant $\tau$, that guarantee robust internal stability of a linear disturbance observer (DOB) under parametric plant uncertainties. Classical small-$\tau$ stability results provide a rigorous foundation for Q-filter bandwidth design. Building on this foundation, we formulate the $\tau$-dependent robust internal-stability problem as a semialgebraic feasibility problem in bivariate polynomials and prove that the complete admissible set is a finite union of open intervals under the standing assumptions. A constructive procedure with exact membership tests recovers all stabilizing intervals. A numerical example with an uncertain plant family containing non-minimum-phase members illustrates a disconnected admissible set.

\end{abstract}

\end{frontmatter}
% Modest extra shrink; nominal spaces unchanged.
\setlength{\parskip}{1pc plus 2pt minus 4pt}
\setlength{\eqntopsep}{12pt plus 2pt minus 4pt}
\setlength{\abovedisplayskip}{12pt plus 2pt minus 4pt}
\setlength{\belowdisplayskip}{12pt plus 2pt minus 4pt}

\section{Introduction}

The disturbance observer (DOB) is widely used for disturbance attenuation and nominal-performance recovery, and its robust stability depends strongly on the Q-filter time constant \(\tau>0\)~\citep{Ohni87,SariyildizOhnishi14,SariyildizOboeOhnishi20}. Classical theory provides small-\(\tau\) sufficient conditions~\citep{ShimJona09,ShimParkJooBackJo16}. \citet{ChanKimH20} established exact vertex and pairwise conditions and organized them into admissible ranges whose intersection identifies robustly stabilizing Q-filter time constants, including for uncertain plant families with non-minimum-phase members. This range-based framework provides the exact robust-stability foundation used here.

Building on this foundation, this paper computes the complete finite structure of the admissible set. It expresses the pairwise condition for the \(\tau\)-dependent DOB characteristic polytope through scalar bivariate polynomial relations, classifies all stability-boundary mechanisms, proves that the set is open and semialgebraic with finitely many connected components, and constructs an exact algebraic decomposition that recovers every component, including disconnected and degenerate cases.

\section{Problem Formulation}\label{sec:problem}
Throughout the paper, $s$ denotes the Laplace variable. Let $P(s)$, $P_\nom(s)$, $C(s)$, and $Q(s;\tau)$ denote the plant, nominal model, outer-loop controller, and Q-filter, respectively, and let $r$, $e$, $u$, $y$, and $d$ denote the reference, tracking error, plant input, output, and disturbance.

\tikzstyle{cyansum} = [draw, fill=myblue, circle, node distance=1cm]
\tikzstyle{cyanblock} = [draw, fill=myblue, rectangle, minimum height=2em, minimum width=3em]
\tikzstyle{block} = [draw, fill=white, rectangle, 
    minimum height=2em, minimum width=3em]
\tikzstyle{sum} = [draw, fill=white, circle, node distance=1cm]

\begin{figure}[!h]
	\centering
	\begin{tikzpicture}[auto, node distance=1.5cm,>=latex']
		\fill[myblue] (2.4,0.4) rectangle (4.25,-2.75);
		\fill[myblue] (2.4,-1.25) rectangle (7.4,-2.75);
		
		\node[coordinate, name = referenceInput]{};
		\node[sum, right of = referenceInput, node distance = 0.5cm, name = sumOuter]{};
		\node[block, right of = sumOuter, node distance = 1.25cm, name = controllerOuter]{$C(s)$};
		\node[cyansum, right of = controllerOuter, node distance = 1cm, name = sumInner]{};
		
		\node[sum, right of = sumInner, node distance = 1.75cm, name = sumDist]{};
		\node[block, right of = sumDist, node distance = 1.5cm, name = Pd]{$P(s)$};

		\node[coordinate, above of = sumDist, node distance = 0.75cm, name = disturbance]{};
		
		\node[coordinate, right of = Pd, node distance = 2cm, name = output]{};
		\node[coordinate, right of = Pd, node distance = 1.6cm, name = outputJunc]{};

		\node[coordinate, right of = sumInner, node distance = 0.75cm, name = dobJunc]{};
		
		\node[cyanblock, below of = Pd, node distance = 2cm, name = combinedInv, minimum width=2.5cm] {$P_\nom^{-1}(s)Q(s;\tau)$};
		\node[coordinate, right of = combinedInv, node distance = 1.6cm, name=temp]{};
		\node[coordinate, below of = temp, node distance = 1cm, name=temp1]{};
        
		\node[cyanblock, below of = dobJunc, node distance = 1cm, name = QB]{$Q(s;\tau)$};
		\node[cyansum, below of = QB, node distance = 1cm, name = sumQ]{};
		
		\draw[->] (referenceInput) -- (sumOuter){} node[pos = 0.5]{$r$};
		\draw[->] (sumOuter) -- (controllerOuter){}node[pos = 0.5]{$e$};
		\draw[->] (controllerOuter) -- (sumInner){};
		\draw[->] (sumInner) -- (sumDist){};
		\draw[->] (sumDist) -- (Pd) node[pos=0.5] {$u$};
		\draw[->] (Pd) -- (output) node[pos = 0.8]{$y$};
		
		\draw[->] (outputJunc) |- (combinedInv){};
		\draw[->] (combinedInv) -- (sumQ);
		
		\draw[->] (dobJunc) -- (QB){};
		\draw[->] (QB) -- (sumQ) node[pos = 0.8, left]{$-$};
		\draw[->] (sumQ) -| (sumInner) node[pos = 0.92]{$-$} node[pos = 0.5]{};
		\draw[->] (disturbance) -- (sumDist) node[pos = 0.2]{$d$};
		\draw[-] (temp) -- (temp1){};
		\draw[->] (temp1) -| (sumOuter) node[pos = 0.96]{$-$} ;
	\end{tikzpicture}
	\caption{Closed-loop system controlled by the inner-loop DOB (shaded block).}
	\label{fig:dob}
\end{figure}
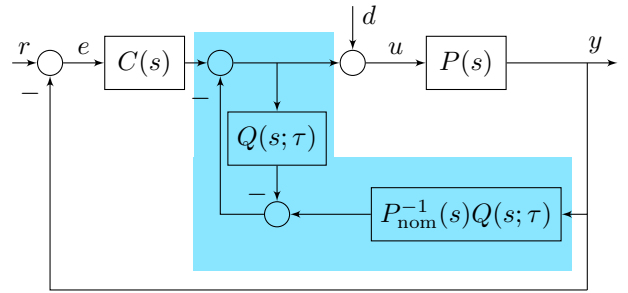

The general configuration of the DOB is depicted in Fig.~\ref{fig:dob}. The actual plant $P(s)\in\Pcal$ is subject to parametric uncertainties, where
\begin{equation}\label{eq:Pcal}
\begin{split}
\Pcal:=\Bigg\{&\frac{\beta_{n-\nu}s^{n-\nu}+\cdots+\beta_1s+\beta_0}
{\alpha_n s^n+\cdots+\alpha_1s+\alpha_0}: \\
&~~~~~~~~\alpha_i\in[\underline{\alpha}_i,\bar{\alpha}_i],\ i=0,\ldots,n,\\
&~~~~~~~~\beta_j\in[\underline{\beta}_j,\bar{\beta}_j],\ j=0,\ldots,n-\nu\Bigg\}.
\end{split}
\end{equation}
Here $n$ and $\nu$ are positive integers satisfying $n\ge\nu$, $\nu$ is the common relative degree, and the interval endpoints in \eqref{eq:Pcal} are known constants. We assume that for every $P(s)\in\Pcal$, its numerator and denominator are coprime and have fixed degrees $n-\nu$ and $n$, respectively.
The block $P_\nom(s)\in\Pcal$ represents the nominal model of the uncertain system $P(s)$, and $C(s)$ denotes a proper outer-loop controller pre-designed for $P_\nom(s)$. The Q-filter is written as
\begin{equation}\label{eq:Q}
Q(s;\tau)=\frac{c_k(\tau s)^k+\cdots+c_1(\tau s)+c_0}
{(\tau s)^l+a_{l-1}(\tau s)^{l-1}+\cdots+a_1(\tau s)+a_0},
\end{equation}
where $l,k\in\ZZ_{\ge0}$, $l\ge k$, $a_i\in\RR$ for $i=0,\ldots,l-1$, and $c_i\in\RR$ for $i=0,\ldots,k$ are fixed coefficients with $c_k\neq0$. We assume that $Q(s;\tau)$ satisfies the following: its denominator polynomial is Hurwitz for all $\tau>0$; it is low-pass with $l-k\geq\nu$, so that the implemented block $P_\nom^{-1}Q$ is proper and is realized without any unstable pole-zero cancellations; and $a_0=c_0\neq0$, so that $Q(0;\tau)=1$. The time constant $\tau>0$ is the key parameter that determines stability and performance of the DOB.

For each $P\in\Pcal$, let $N$ and $D$ be the numerator and denominator in \eqref{eq:Pcal}, and fix coprime representations of the remaining blocks, with arguments suppressed:
\[
P_\nom=\frac{N_\nom}{D_\nom},\qquad
C=\frac{N_C}{D_C},\qquad
Q=\frac{N_Q}{D_Q}.
\]
Arguments are restored whenever needed.
Then, the transfer function matrix from $[r(s),d(s)]^\top$ to $[e(s),u(s)]^\top$ in Fig.~\ref{fig:dob} is expressed as
\begin{equation}\label{eq:TM}
\frac{1}{\Lambda}\begin{bmatrix}
    Q(P-P_\nom) + P_\nom & -P_\nom P(1-Q) \\
    P_\nom C & P_\nom(1-Q)
\end{bmatrix},    
\end{equation}
where
$\Lambda:= P_\nom(1+PC)+Q(P-P_\nom)$. Equation~\eqref{eq:TM} gives the closed-loop transfer matrix associated with the reference and disturbance inputs. The implemented interconnection in Fig.~\ref{fig:dob} is internally stable if it is well posed and all closed-loop transfer functions from independent exogenous inputs injected at the internal summing junctions to all internal signals are proper and stable. Hidden unstable modes and unstable pole--zero cancellations are not regarded as internally stable. The closed-loop system is robustly internally stable if this property holds for every $P(s)\in\Pcal$~\citep{ZhouDoyl98}.

The following proposition, recalled from \citet[Theorem~3, sufficiency]{ShimJona09}, states sufficient small-$\tau$ conditions for robust internal stability.

\begin{proposition}
\label{prop:tau*}
There exists a constant $\tau^*>0$ such that for all $0<\tau\leq \tau^*$, the closed-loop system in Fig.~\ref{fig:dob} is robustly internally stable if the following conditions hold:
\begin{enumerate}
    \item the nominal feedback interconnection of $P_\nom(s)$ and $C(s)$ is internally stable,
    \item $P(s)$ is minimum-phase for all $P(s)\in\Pcal$,
    \item the polynomial
    \[
    D_Q(s;1)+
    \left(
    \lim_{s\to\infty}\frac{P(s)}{P_\nom(s)}-1
    \right)N_Q(s;1)
    \]
    is Hurwitz for all $P(s)\in\Pcal$.
\end{enumerate}
\end{proposition}

The following characteristic-polynomial test is recalled from \citet[Lemma~2]{ChanKimH20}.

\begin{lemma}
\label{lem:delta}
    The closed-loop system in Fig.~\ref{fig:dob} is robustly internally stable if and only if the plant-dependent characteristic polynomial
    \begin{equation}\label{eq:delta}
    \begin{split}
    \delta(s;\tau;P)
    &=(DD_C+NN_C)N_\nom D_Q \\
    &\quad+N_QD_C(ND_\nom-N_\nom D)
    \end{split}
    \end{equation}
    is Hurwitz for all $P(s)\in\Pcal$.
\end{lemma}

\begin{remark}\label{rem:nmp}
Because \(P_\nom\in\Pcal\), a nonempty admissible set requires \(N_\nom\) to have no zeros outside the open left-half plane under Lemma~\ref{lem:delta}; hence non-minimum-phase plants may occur only among the nonnominal members of \(\Pcal\), not in \(P_\nom\).
\end{remark}

Define
\[
\begin{aligned}
p_D&:=D_CN_\nom(D_Q-N_Q),\\
p_N&:=N_CN_\nom D_Q+N_QD_CD_\nom.
\end{aligned}
\]
Then \eqref{eq:delta} becomes $\delta(s;\tau;P)=Dp_D+Np_N$. Its affine dependence on the uncertain coefficients gives, for each $\tau>0$,
\[
\{\delta(\cdot;\tau;P):P\in\Pcal\}=\operatorname{co}\Delta(\tau)\subset\RR[s],
\]
where $\operatorname{co}$ denotes convex hull and
\begin{equation}\label{eq:Delta}
\begin{split}
\Delta(\tau):= 
\bigg\{ &\Big(\sum_{i=0}^{n}v_i s^i\Big) p_D(s;\tau) + \Big(\sum_{j=0}^{n-\nu}w_j s^j\Big) p_N(s;\tau):\\ 
&v_i\in \{\underline{\alpha}_i, \bar{\alpha}_i\},~ i=0,\ldots,n, \\
&w_j\in\{\underline{\beta}_j, \bar{\beta}_j\},~ j=0,\ldots,n-\nu  \bigg\}.
\end{split}    
\end{equation}
After removing corner polynomial families that are identical as elements of $\RR[s,\tau]$, enumerate the remaining families as $\delta_i(s;\tau)$, $i=1,\ldots,m$, where $m\leq 2^{2n-\nu+2}$.

\begin{lemma}\label{lem:degree}
Under the standing assumptions, all corner polynomials have degree $\rho:=2n-\nu+\deg D_C+l$ and a common nonzero leading sign. The fixed coprime representation of $P_\nom$ can be chosen so that this sign is positive.
\end{lemma}
\begin{proof}
The term $DD_CN_\nom D_Q$ in~\eqref{eq:delta} has degree $\rho$. Properness of $C$ and $l-k\geq\nu$ make every other term have degree at most $\rho-\nu$, so cancellation at degree $\rho$ is impossible. Writing $\operatorname{lc}$ for leading coefficient, the coefficient of $s^\rho$ is $v_n\operatorname{lc}(D_C)\operatorname{lc}(N_\nom)\tau^l$. The fixed denominator degree implies $0\notin[\underline{\alpha}_n,\bar{\alpha}_n]$, so this sign is common and nonzero. Reversing the common signs of $N_\nom,D_\nom$, if necessary, preserves $P_\nom$ and all roots while making the leading sign positive.
\end{proof}

For each $(i,j)\in\Ical:=\{(i,j)\in\ZZ^2:1\le i<j\le m\}$, define the pairwise \textit{segment polynomial}
\[
p_{ij}(s;\tau;\lambda)
:=\lambda\delta_i(s;\tau)+(1-\lambda)\delta_j(s;\tau),\qquad
\lambda\in[0,1].
\]
Although not every pairwise segment is exposed, testing all pairs is exact because every tested segment lies in $\operatorname{co}\Delta(\tau)$ and every exposed edge is tested. The following all-pairs form of the Edge theorem~\citep{BartHoll88} is recalled from \citet[Lemma~3]{ChanKimH20}.

\begin{lemma}
\label{lem:edge}
The closed-loop system in Fig.~\ref{fig:dob} is robustly internally stable if and only if every corner polynomial is Hurwitz and, for all $(i,j)\in\Ical$ and $\lambda\in[0,1]$, the segment polynomial $p_{ij}(s;\tau;\lambda)$ is Hurwitz.
\end{lemma}

The vertex and pairwise ranges of \citet{ChanKimH20} provide the exact admissible-range framework used below. The present analysis equips this framework with a complete finite structure by classifying all finite boundary mechanisms and constructing an exact algebraic decomposition that recovers every connected component.

The problem addressed in this paper is formally stated as follows.

\begin{problem}
Given $C(s)$, $P_\nom(s)$, $Q(s;1)$, and the set of uncertain plants $\Pcal$ in \eqref{eq:Pcal}, find the entire set of admissible $\tau>0$ such that the closed-loop system in Fig.~\ref{fig:dob} is robustly internally stable.  
\end{problem}

\section{Main Result}\label{sec:main}

In this section, we reformulate the conditions of Lemma~\ref{lem:edge} into a set of equations and inequalities involving bivariate polynomials. Subsequently, we demonstrate that the set of admissible $\tau$ is a finite union of open intervals. Finally, we propose an algorithm to determine the endpoints of these intervals.

\subsection{Reformulation of Stability Condition}
For any $i\in\{1,\ldots,m\}$, let $\mathrm{i}:=\sqrt{-1}$, take $\omega\in\RR$, and set $\eta:=\omega^2\geq0$. Since the coefficients of $\delta_i(s;\tau)$ are real, define the real bivariate polynomials $A_i$ and $B_i$ by
\begin{equation}\label{eq:AB}
\delta_i(\mathrm{i}\omega; \tau) =: A_i(\eta; \tau) + \mathrm{i}\omega B_i(\eta;\tau).    
\end{equation}
For any $(i,j)\in\Ical$, let us also define
\begin{equation}\label{eq:GF}
\begin{split}
G_{ij}(\eta;\tau) &:= A_i(\eta;\tau) B_j(\eta;\tau) - A_j(\eta;\tau) B_i(\eta;\tau), \\
    F_{ij}(\eta;\tau) &:= A_i(\eta;\tau) A_j(\eta;\tau) + \eta B_i(\eta;\tau) B_j(\eta;\tau).
\end{split}
\end{equation}

The following proposition expresses the classical fixed-\(\tau\) segment-stability condition~\citep[Chapter~2]{BhattacharyyaChapellatKeel95} in the bivariate form needed for the \(\tau\)-dependent DOB segment family.

\begin{proposition}\label{prop:equiv}
Fix $\tau>0$ and $(i,j)\in\Ical$. Assume that both $\delta_i(s;\tau)$ and $\delta_j(s;\tau)$ are Hurwitz.  Then, the following two statements are equivalent.
\begin{enumerate}
\item[(i)] There exists $\lambda^*\in[0,1]$ such that $p_{ij}(s;\tau;\lambda^*)$ is not Hurwitz.
\item[(ii)] There exists $\eta^*>0$ such that
\begin{equation}\label{eq:GFcon}
    G_{ij}(\eta^*;\tau)=0,~ F_{ij}(\eta^*;\tau) \leq0.
\end{equation}
\end{enumerate}
\end{proposition}
\begin{proof}
Suppose that (i) holds true. Lemma~\ref{lem:degree} excludes degree loss and roots escaping through infinity along the segment family. Hence, by continuity of roots in $\lambda$, there exist $\lambda_0\in(0,1)$ and $\omega_0\in\RR$ such that
\begin{equation}\label{eq:temp1}
p_{ij}(\mathrm{i}\omega_0;\tau;\lambda_0) = \lambda_0 \delta_i(\mathrm{i}\omega_0;\tau) + (1-\lambda_0) \delta_j(\mathrm{i}\omega_0;\tau) =0    
\end{equation}
because both endpoint polynomials $\delta_i(s;\tau)$ and $\delta_j(s;\tau)$ are Hurwitz.
Because both endpoint polynomials are Hurwitz with positive leading coefficients, their constant terms are positive; hence the constant term of $p_{ij}(s;\tau;\lambda)$ is positive for every $\lambda\in[0,1]$, so $\omega_0\neq0$. Set $\eta^*:={\omega_0}^2>0$. Since both endpoint polynomials are Hurwitz, $\delta_i(\mathrm{i}\omega_0;\tau)$ and $\delta_j(\mathrm{i}\omega_0;\tau)$ are nonzero; by \eqref{eq:temp1}, they are collinear and oppositely directed. By \eqref{eq:AB}, collinearity and $\omega_0\neq0$ imply $G_{ij}(\eta^*;\tau)=0$, while the opposite direction implies $F_{ij}(\eta^*;\tau)\leq0$. Therefore, (ii) holds true.

Conversely, suppose that there exists $\eta^*>0$ such that $G_{ij}(\eta^*;\tau)=0$ and $F_{ij}(\eta^*;\tau)\leq0$. Then, the complex numbers $\delta_i(\mathrm{i}\omega_0;\tau)$ and $\delta_j(\mathrm{i}\omega_0;\tau)$ with 
$\omega_0=\sqrt{\eta^*}$ are collinear and oppositely directed. Therefore, there exists $\lambda^*\in[0,1]$ such that $p_{ij}(s;\tau;\lambda^*)$ has a pure imaginary root $s=\mathrm{i}\omega_0$ so that it is not Hurwitz, which concludes the proof.
\end{proof}

Proposition~\ref{prop:equiv} provides the bridge from fixed-\(\tau\) segment tests to global DOB bandwidth analysis. Its joint \((\eta,\tau)\) representation makes the segment-boundary mechanisms algebraically accessible and supports the exact admissible-set construction developed below.

\subsection{Construction of the Set of Admissible $\tau$}\label{sec:steps}

For each $i=1,\ldots,m$, define
\begin{equation}\label{eq:V}
V_i:= \{\tau>0: \delta_i(s;\tau)\text{ is Hurwitz} \}.    
\end{equation}
Also, for each $(i,j)\in\Ical$, let us define
\begin{equation}\label{eq:E}
\begin{aligned}
E_{ij}:=\bigl\{\tau>0:\ \nexists\,\eta>0\ \text{ with }&
G_{ij}(\eta;\tau)=0,\\[-1mm]
&F_{ij}(\eta;\tau)\leq0\bigr\}.
\end{aligned}
\end{equation}
Then, according to Lemma~\ref{lem:edge} and Proposition~\ref{prop:equiv}, the exact admissible set is
\begin{equation}\label{eq:Rcal}
\Rcal:= \big( \bigcap_{i=1}^m V_i\big) \cap \big( \bigcap_{(i,j)\in\Ical} E_{ij} \big),
\end{equation}
with the second intersection understood as $(0,\infty)$ when $\Ical=\emptyset$. For each $i\in\{1,\ldots,m\}$, let $H_i(\tau)$ be the Hurwitz matrix of $\delta_i(s;\tau)$, and let $M_i^q(\tau)$, $q=1,\ldots,\rho$, denote the $q$-th leading principal minor of $H_i(\tau)$. If $M_i^q\equiv0$ for some $i,q$, then $V_i=\emptyset$ and hence $\Rcal=\emptyset$; below we consider the nontrivial case in which all $M_i^q$ are nonzero.

\begin{theorem}\label{thm:semi}
The admissible set $\Rcal$ in \eqref{eq:Rcal} is an open semialgebraic subset of $(0,\infty)$. Hence it is a finite union of possibly unbounded open intervals.
\end{theorem}
\begin{proof}
Hurwitz stability of each corner polynomial is expressed by the Routh--Hurwitz inequalities \(M_i^q(\tau)>0\), \(q=1,\ldots,\rho\); hence these conditions are semialgebraic in \(\tau\). By Proposition~\ref{prop:equiv}, the relative complement of each pairwise segment condition is the projection of the semialgebraic set
\[
\eta>0,\quad \tau>0,\quad
G_{ij}(\eta;\tau)=0,\quad F_{ij}(\eta;\tau)\le0.
\]
The Tarski--Seidenberg theorem and closure of semialgebraic sets under finite Boolean operations imply that \(\Rcal\) is semialgebraic~\citep{BasuPollackRoy06}. To prove openness, fix \(\tau_0\in\Rcal\). The coefficient images of the finitely many corner polynomials and of all pairwise segment polynomials \(p_{ij}(s;\tau_0;\lambda)\), \(\lambda\in[0,1]\), form compact subsets of the open Hurwitz region. By joint continuity in \((\tau,\lambda)\), compactness of \([0,1]\), and finiteness of the corner pairs, the same remains true for all \(\tau\) in a neighborhood of \(\tau_0\). Hence \(\Rcal\) is open. The final claim follows from the structure theorem for semialgebraic subsets of \(\RR\)~\citep{BasuPollackRoy06}.
\end{proof}

Theorem~\ref{thm:semi} reduces the construction of \(\Rcal\) to locating the boundary points of its open components. The following theorem gives finite algebraic necessary conditions for such boundary points.

\begin{theorem}\label{thm:main}
Suppose that $\tau^\circ>0$ is an endpoint of a connected component of $\Rcal$. Then, at least one of the following holds.
\begin{enumerate}
\item[(i)] There exist $i\in\{1,\ldots,m\}$ and $q\in\{1,\ldots,\rho\}$ such that
\begin{equation}\label{eq:i}
M_i^q(\tau^\circ)=0.    
\end{equation}
\item[(ii)] There exist $(i,j)\in\Ical$ and $\eta^\circ >0 $ such that 
\begin{equation}\label{eq:ii}
G_{ij}(\eta^\circ;\tau^\circ)=0,~F_{ij}(\eta^\circ;\tau^\circ)=0.
\end{equation}
\item[(iii)] There exist $(i,j)\in\Ical$ and $\eta^\circ>0$ such that 
\begin{equation}\label{eq:iii}
\begin{split}
G_{ij}(\eta^\circ;\tau^\circ)&=0,~F_{ij}(\eta^\circ;\tau^\circ)\leq0, \\
\partial_\eta G_{ij}(\eta^\circ;\tau^\circ)&=0.    
\end{split}
\end{equation}
\end{enumerate}
\end{theorem}
\begin{proof}
Since $\Rcal$ is open, $\tau^\circ\notin\Rcal$. Because $\tau^\circ$ is an endpoint of a connected component of $\Rcal$, there exists a sequence $\{\widehat{\tau}_\ell\}\subset\Rcal$ such that $\widehat{\tau}_\ell\to\tau^\circ$.

First, suppose that a corner polynomial $\delta_i(s;\tau^\circ)$ is not Hurwitz. Since $\widehat{\tau}_\ell\in\Rcal$,
\[
M_i^q(\widehat{\tau}_\ell)>0,\qquad q=1,\ldots,\rho.
\]
By continuity, $M_i^q(\tau^\circ)\geq0$ for every $q$. If all these inequalities were strict, the Routh--Hurwitz criterion would imply that $\delta_i(s;\tau^\circ)$ is Hurwitz. Hence $M_i^q(\tau^\circ)=0$ for at least one $q$, and condition~(i) holds.

Otherwise, all corner polynomials are Hurwitz at $\tau^\circ$, and some pairwise segment condition must fail. Therefore, there exist $(i,j)\in\Ical$ and $\eta^\circ>0$ such that
\[
G_{ij}(\eta^\circ;\tau^\circ)=0,\qquad
F_{ij}(\eta^\circ;\tau^\circ)\leq0.
\]
Suppose that neither condition~(ii) nor condition~(iii) holds. Then
\[
F_{ij}(\eta^\circ;\tau^\circ)<0,\qquad
\partial_\eta G_{ij}(\eta^\circ;\tau^\circ)\neq0.
\]
The implicit function theorem provides a differentiable map $\phi$ such that
\[
\phi(\tau^\circ)=\eta^\circ,\qquad
G_{ij}(\phi(\tau);\tau)=0
\]
in a neighborhood of $\tau^\circ$. By reducing the neighborhood if necessary, we may ensure that $\phi(\tau)>0$ and $\tau>0$ throughout it. By continuity of $F_{ij}$, there exists $\epsilon>0$ such that
\[
F_{ij}(\phi(\tau);\tau)<0,\qquad
G_{ij}(\phi(\tau);\tau)=0
\]
for every $\tau\in(\tau^\circ-\epsilon,\tau^\circ+\epsilon)$. Consequently,
\[
(\tau^\circ-\epsilon,\tau^\circ+\epsilon)\cap E_{ij}=\emptyset,
\]
and hence this neighborhood is disjoint from $\Rcal$. This contradicts the existence of the sequence $\widehat{\tau}_\ell\in\Rcal$ converging to $\tau^\circ$. Therefore, at least one of conditions~(i)--(iii) holds.
\end{proof}

Theorem~\ref{thm:main} supplies interpretable boundary mechanisms, whereas quantifier elimination guarantees completeness in degenerate cases. The characterization~\eqref{eq:Rcal} is exact set-theoretically without a genericity assumption. For rational or real-algebraic input data, real quantifier elimination and algebraic sign determination give a finite exact implementation. Specifically, eliminating \(\eta\)~\citep{BasuPollackRoy06} yields a quantifier-free formula \(\Phi\) such that \(\tau\in\Rcal\Longleftrightarrow\Phi(\tau)\) for \(\tau>0\), with finitely many sign conditions on univariate polynomials. Ordinary resultants provide convenient elimination polynomials for conditions~\eqref{eq:ii}--\eqref{eq:iii} when nonzero, while quantifier elimination covers degenerate cases.
\\
\textbf{Step 1:}
After simplifying identically zero polynomials in \(\Phi\), let \(\mathcal C\subset(0,\infty)\) be the set of positive real roots of all nonzero polynomials appearing in \(\Phi\). Then \(\partial\Rcal\cap(0,\infty)\subseteq\mathcal C\), where \(\partial\Rcal\) denotes the boundary in \(\RR\). If nonempty, enumerate \(\mathcal C\) as \(0<\xi_1<\cdots<\xi_K\).
\\
\textbf{Step 2:}
If \(\mathcal C=\emptyset\), evaluate the sole open cell \((0,\infty)\). Otherwise, evaluate \(\Phi\) at one sample point from each open cell \((0,\xi_1),(\xi_1,\xi_2),\ldots,(\xi_K,\infty)\) and at every singleton \(\{\xi_h\}\), \(h=1,\ldots,K\). Polynomial signs are constant on each open cell, so every such cell is either contained in \(\Rcal\) or disjoint from it.
\\
\textbf{Step 3:}
Take the union of all admissible open and zero-dimensional cells, and merge each maximal consecutive collection whose union is connected. Since \(\Rcal\) is open, the result is its exact decomposition into open intervals. Equivalently, fixed-\(\tau\) membership is decided by the corner Hurwitz tests and the nonexistence, for every \((i,j)\in\Ical\), of \(\eta>0\) satisfying \(G_{ij}=0\) and \(F_{ij}\leq0\). The exact set \(\Rcal\) then serves as the robust-stability domain for bandwidth tuning.

\section{Numerical Example}\label{sec:ex}

Consider
\begin{align*}
C(s)&=\frac{1}{s+1},\qquad Q(s;\tau)=\frac{3\tau s+1}{(\tau s)^2+8\tau s+1},\\
P_{\beta_1}(s)&=\frac{s^2+\beta_1s+5}{s^3+3s^2+3s+1},\qquad \beta_1\in[-0.2,2].
\end{align*}
Let \(P_\nom(s)=\left.P_{\beta_1}(s)\right|_{\beta_1=1}\). This family contains a non-minimum-phase plant and has two degree-eight corner polynomials. Let \(\mathcal C_V\), \(\mathcal C_{GF}\), and \(\mathcal C_{GG_\eta}\) be the exact positive-root sets of the corner Hurwitz minors, \(\Res_\eta(G_{12},F_{12})\), and \(\Res_\eta(G_{12},\partial_\eta G_{12})\), respectively, where \(\Res_\eta\) denotes the resultant in \(\eta\). Their decimal approximations are
\begin{align*}
\mathcal C_V\approx{}&\{0.028,0.033,0.035,0.232,0.489,\\[-1mm]
&\hspace{8mm}1.748,5.204,410.649\},\\
\mathcal C_{GF}\approx{}&\{0.035,0.127,0.232,0.489,\\[-1mm]
&\hspace{8mm}1.748,7.873\},\\
\mathcal C_{GG_\eta}\approx{}&\{0.077,2.662,3.655,6.525\}.
\end{align*}
Use \(\widetilde{\mathcal C}:=\mathcal C_V\cup\mathcal C_{GF}\cup\mathcal C_{GG_\eta}\) as the boundary-containing set in Steps~2--3. Exact duplicate removal gives 14 points, and the cell tests resolve extraneous roots; decimals are shown only for readability. Let \(0<\tau_1^\star<\tau_2^\star<\tau_3^\star<\tau_4^\star\) be the exact boundary points returned by the classification. Then
\[
\Rcal=(\tau_1^\star,\tau_2^\star)\cup(\tau_3^\star,\tau_4^\star),
\]
where
\begin{align*}
(\tau_1^\star,\tau_2^\star)&\approx(0.0345520,0.2321333),\\
(\tau_3^\star,\tau_4^\star)&\approx(0.4885786,1.7483813).
\end{align*}
Thus \(\Rcal\) is disconnected and contains an admissible component away from the origin, illustrating a global alternative to local small-\(\tau\) bandwidth selection.

\bibliographystyle{agsm}
\bibliography{Hamin1}
\end{document}